\documentclass[final]{ias2}

\usepackage{graphicx} 
\usepackage{multirow}
\usepackage{array} 

\usepackage{hyperref} 
\def\Journal#1#2#3#4{{#1} {\bf #2},{#3}{(#4)}}

\def\NPB{{\em Nucl. Phys.} {\bf B}}

\def\NPA{{\em Nucl. Phys.} A}
\def\MPLA{\em Mod. Phys. Lett. {\bf A}}
\def\PLB{{\em Phys. Lett.}  {\bf B}}
\def\PRL{\em Phys. Rev. Lett.}
\def\PRD{{\em Phys. Rev.} {\bf D}}
\def\PRC{{\em Phys. Rev.} C}

\def\ZPA{{\em Z. Phys.} A}

\def\JINST{\em JINST}

\def\RPP{\em Rep. Prog. Phys.}

\def\IJMP{\em Int. Journal of Modern Physics E}
\def\JPG{\em Journal of Physics G}
\def\NJP{\em New Journ. Phys.}

\def\ra{\rightarrow}
\def\be{\begin{equation}}
\def\ee{\end{equation}}

\newcommand{\nme}{nuclear matrix element }

\newcommand{\bpbp}{\mbox{$\beta^+\beta^+$} }
\newcommand{\ecec}{\mbox{$EC/EC$} }
\newcommand{\bec}{\mbox{$\beta^+/EC$} }

\newcommand{\bnel}{\mbox{$\bar{\nu}_e$} }

\newcommand{\nel}{\mbox{$\nu_e$}}

\newcommand{\ton}{\mbox{$T_{1/2}^{0\nu}$} }
\newcommand{\tzn}{\mbox{$T_{1/2}^{2\nu}$} }

\newcommand{\ema}{\mbox{$\langle m_{ee} \rangle$ }}
\newcommand{\bea}{\begin{equation} \begin{array}{c}}
\newcommand{\eea}{ \end{array} \end{equation}}

\newcommand{\gess}{\mbox{$^{76}$Ge }}
\newcommand{\seza}{\mbox{$^{82}$Se }} 
\newcommand{\caav}{\mbox{$^{48}$Ca }}

\newcommand{\cd}{\mbox{$^{116}$Cd }}

\newcommand{\nd}{\mbox{$^{150}$Nd} }

\newcommand{\tehd}{\mbox{$^{130}$Te }}
\newcommand{\xehs}{\mbox{$^{136}$Xe }}

\newcommand{\obb}{0\mbox{$\nu\beta\beta$-decay }}

\newcommand{\zbb}{2\mbox{$\nu\beta\beta$-decay }}

\begin{document}

\markboth{Kai Zuber}{Kai Zuber}

\title{Neutrinoless double beta decay}

\author[sin]{Kai Zuber} 
\email{zuber@physik.tu-dresden.de}
\address[sin]{Institute for Nuclear and Particle Physics, TU Dresden, Germany}

\begin{abstract}
The physics potential of neutrinoless double beta decay is discussed. Furthermore,
experimental considerations are presented as well as the current status of experiments.
Finally an outlook towards the future, work on nuclear matrix elements and alternative
processes is given.
\end{abstract}


 
\maketitle


\section{Introduction and physics}
\label{sec:intro}
Neutrinos play a crucial role in modern particle, nuclear and astrophysics including cosmology. 
It has been the major achievement of the last 15 years to show that neutrinos have a non-vanishing 
rest mass. The evidence arises from a deficit of upward going atmospheric muon neutrinos
confirmed by long baseline accelerator experiments and the solution of 
the solar neutrino problem being confirmed by nuclear reactors. All can be explained 
by neutrino oscillations, which are depending mass differences $\Delta m^2_{ij} = m_j^2 - m_i^2$
assuming a two flavour mixing only. 
The determination of absolute neutrino masses is now a major issue, because neutrino oscillation experiments do not
allow this. The classical way search for a rest mass of the neutrino is the study of the endpoint region of electron
spectra in beta decay (see \cite{ott08} for a recent review). The KATRIN experiment is well on its way
to improve the current bound of about 2.2 eV for \bnel~ by an order of magnitude. 
Further bounds on the total sum of neutrino masses can be obtained from cosmological
studies.
Another laboratory process is the rare nuclear decay of neutrinoless double beta decay.
\be
\label{eq:0nu}
(Z,A) \ra (Z+2,A) + 2 e^-  \quad(\obb)
\ee
Single beta decay  must be forbidden or at least strongly suppressed to observe this
decay and 35 potential isotopes exist in nature.
As can be seen from Equation \ref{eq:0nu} the given decay mode is violating total lepton number by two units 
and thus not allowed in the
Standard Model. Being a decay the observable is a half-life which can be linked to the
 quantity of interest $\epsilon$ via  
\be
\label{eq:conv}
(\ton )^{-1} = G_{PS}  \mid M_{Nuc} \mid^2 \epsilon^2
\ee
with $G_{PS}$ being the phase space, $\mid M_{Nuc} \mid$ 
the involved nuclear matrix element for the physics process considered
to describe this decay and $\epsilon$ the quantity of interest. 
Various BSM processes can be considered among
them light and heavy Majorana neutrino exchange , right-handed weak currents,
R-parity violating SUSY ($\lambda'_{111}$) and double charged Higgs bosons.
If \obb is ever observed it was shown that this would imply that neutrinos
are Majorana particles \cite{sch81}. However, given all the possible processes it 
will become an important question what the individual contributions
of the considered processes will be. The LHC will help to restrict this by
performing searches for new particles in the TeV range. An example how this complementary information
from LHC and \obb can be used within the context of left-right symmetric theories is given in \cite{tel11}.
For a recent extensive review on the particle physics in double decay see \cite{rod11}. \\
The standard interpretation considered here is the one using light Majorana neutrino exchange.
In this case $\epsilon$ is the effective Majorana neutrino mass \ema  given by
\be
\label{eq:ema} \epsilon \equiv \ema = \mid \sum_i U_{ei}^2 m_i\mid
\ee
with $U_{ei}^2$ as the mixing matrix elements containing the electron neutrino. 

\section{Double beta decay and neutrino oscillation results}
\begin{figure}[ht]
\begin{center}
\includegraphics[width=0.8\columnwidth]{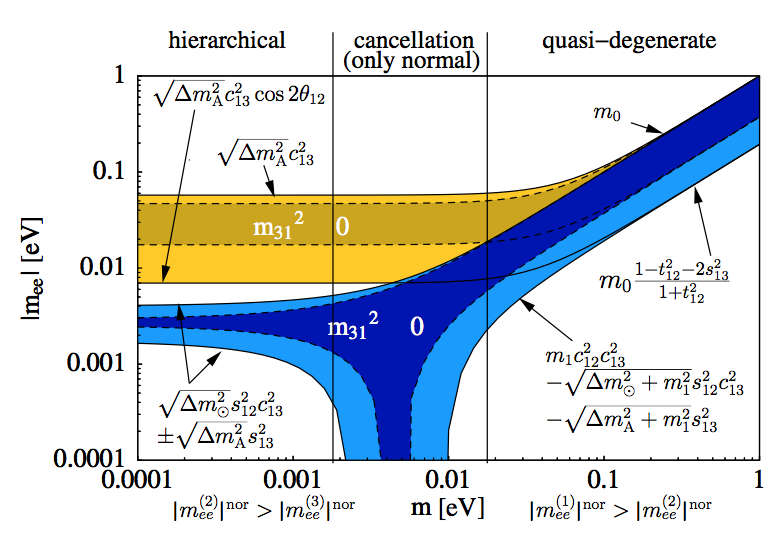}
\caption{The effective Majorana neutrino mass \ema as a function of the smallest neutrino mass and the 
principal dependence of individual features on the oscillation parameters. A value of $\sin^2 2\theta_{13}$ = 0.02
has been assumed, furthermore $t^2_{12} = \tan^2 \theta_{12}$. Uncertainties introduced due to the 
involved nuclear matrix elements are not included (from \cite{lin06}).}
\label{fig:meff}
\end{center}
\end{figure}

For the following discussion a restriction to the light Majorana neutrino case is done.
It is evident from Eq.~\ref{eq:ema} that the expectation for \ema in double beta decay 
depends on the neutrino oscillation parameters. It should be noted that the mixing matrix
of relevance is given by 
\be
U = U_{PMNS} (\theta_{12}, \theta_{13}, \theta_{23}, e^{i\delta}) \times  diag (1, e^{i\alpha}, e^{i\beta})
\ee
with the standard leptonic mixing matrix U$_{PMNS}$ and two additional CP-phases $\alpha$ and
$\beta$, called Majorana phases, which appear if neutrinos are their own antiparticle. These
phases do not show up in oscillation experiments. Hence, \ema can be written 
as a sum of three terms
\be
\ema = \mid m_e^1 \mid + \mid m_e^2 \mid e^{2i\alpha} + \mid m_e^3 \mid  e^{2i\beta}
\ee
with the individual contributions given as
\bea
m_e^1   =   \mid U_{e1} \mid^ 2 m_1 = m_1 \cos^2 \theta_{12} \cos^2\theta_{13} \\
m_e^2   =   \mid U_{e2} \mid^ 2 m_2 = m_2 \sin^2 \theta_{12} \cos^2\theta_{13}  \\
m_e^3   =  \mid U_{e3} \mid^ 2 m_3 = m_3  \sin^2\theta_{13}
\eea
Latest global fits to available oscillation parameters are given in \cite{sch11,fog11}. 
An important new ingredient is that 3-flavour fits to solar neutrinos \cite{aha11}, observations from T2K \cite{abe11} and MINOS 
\cite{ada11}
as well as first Double Chooz results \cite{ker11} favour a non-zero value of $\theta_{13}$.
As the oscillations do not fit the absolute scale there are two options for arranging the
mass eigenstates, either $m_3 > m_2 > m_1$ (normal hierarchy, NH) or $ m_2 > m_1 > m_3$
(inverse hierarchy,IH). If the neutrino masses turn out to be close the the current limit
from beta decay there is a quasidegeneracy ($m_1 \approx m_2 \approx m_3  \equiv m_0$).
In case of hierarchies the two larger masses can be expressed as
\bea
m_2  = \sqrt{m_1^2 + \Delta m_\odot^2} \qquad m_3  =   \sqrt{m_1^2 + \Delta  m_{atm}^2 } \qquad \mbox{(normal)} \\
m_2 = \sqrt{m_3^2 + \Delta m_\odot^2 + \Delta  m_{atm}^2} \qquad m_1 = \sqrt{m_3^2 + \Delta  m_{atm}^2} \qquad \mbox{(inverted)} 
\eea
with the solar splitting $\Delta m_\odot^2 = 7.59^{+0.20}_{-0.18} \times 10^{-5} eV^2 $ and the atmospheric splitting
 $ \Delta  m_{atm}^2 =  2.49 \pm 0.09 \times 10^{-3} eV^2 $ (normal) , 
 $ \Delta  m_{atm}^2 =  - 2.343^{+0.10}_{-0.09} \times 10^{-3} eV^2 $ (inverted)   \cite{sch11}.
The principal behaviour of \ema as a function of the lightest mass eigenstate is shown in Figure~\ref{fig:meff} 
for an exemplaric value
of $\sin^2 2\theta_{13}$ = 0.02 and the dependence of individual parts on the parameters.
For \ema values larger than about 100 meV neutrinos are almost degenerate, the inverted hierarchy covers a range between
about 10-50 meV and below 10 meV is the region of the normal hierachy. As can be seen, in the NH there can be a cancellation
among the terms, the allowed region for that becomes larger with increasing $\sin^2 2\theta_{13}$. There is no such effect in 
the IH because of the non-maximal solar mixing angle $\theta_{12}$. In addition the gap between the two bands in 
the hierarchical region will shrink with increasing $\sin^2 2\theta_{13}$. Half-lives for the IH are in the region beyond 10$^{26}$years
while half-lives in the NH are well beyond 10$^{28}$ years.

\section{General experimental considerations}
Evidently measurements of half-lives well beyond 10$^{25}$ years are by now
means trivial. As signal for the process given in Equation (\ref{eq:0nu}) serves a peak in the sum energy spectrum 
of the two electrons equivalent to the Q-value of the nuclear transition. 
The corresponding half-life in case of no background is given by the radioactive
decay law
\be
\ton = \ln 2 m a t N_A /N_{\beta\beta}
\ee
with $m$ the used mass, $a$ the isotopical abundance of the double beta emitter, $t$ the measuring time, $N_A$ the 
Avogadro constant and $N_{\beta\beta}$ the number of double beta events, which has to be taken from the experiment.
If no peak is observed and a constant background (i.e. all potential energy depositions
in the region of interest, i.e. around the Q-value, not being neutrinoless double beta decay) is assumed scaling
linearly with time, $N_{\beta\beta}$ is derived as
\be
\ton \propto a \times \epsilon \sqrt{\frac{M \times t}{B \times \Delta E}}
\ee
where $\epsilon$ is the efficiency for detection of the total energy of both electrons, 
$\Delta E$ is the energy resolution at the
peak position and $B$ the background index normally given in counts/keV/kg/year.
Hence, the most crucial parameters are a high detection efficiency and high abundance
of the isotope of interest. This is the reason why almost all next generation experiments
are using enriched materials and the ''source = detector'' approach. 
Furthermore, the energy resolution\footnote{Care must be taken as for traditional reasons different detector technologies
use either the Gaussian $\sigma$ or the Full Width at Half Maximum $\Delta E$ to quote energy
resolution. The relation among the quantities is $\Delta E =2.35 \sigma$.} should be
as good as possible to concentrate the few expected events in a small region and
ideally the experiment should be background free. An irreducible background is
the Standard Model process \zbb 
\be
\label{eq:2nu}
(Z,A) \ra (Z+2,A) + 2 e^- + 2 \nel  \quad(\zbb).
\ee
\begin{figure}[ht]
\begin{center}
\includegraphics[width=0.8\columnwidth]{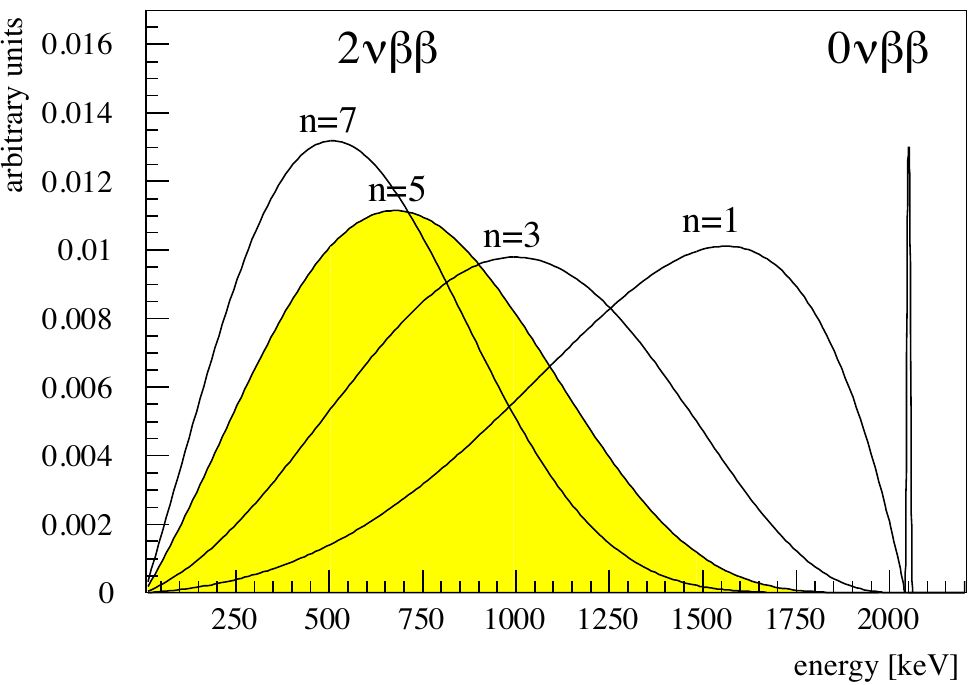}
\caption{Schematic plot of the sum energy spectrum of the two electrons in double beta decay. \obb
results in a peak at the Q-value of the transition. Various modes can be characterised by the phase
space dependence (Q-E)$^n$. The mode n=5 is the \zbb while the modes n=1,3,7 involve the 
emission of a majoron, a Goldstone boson linked to the spontaneous breaking of lepton number.
The different modes belong to different behaviours of the majoron with respect to weak isospin.
The individual contributions are not to scale.}
\label{fig:spectrum}
\end{center}
\end{figure}
Here again energy resolution matters, because of the continuous spectrum of the \zbb
mode it is only its high energy part leaking into the peak region (see Figure~(\ref{fig:spectrum})).
Nevertheless, this can be a worry as the half-life is typically several orders of magnitude
shorter than the expected  one for \obb.
As the decay rate for \obb~ scales with Q$^5$ only isotopes with Q-values
above 2 MeV are considered for experimental searches. They are listed together with some
important numbers in Table~(\ref{tab:isotopes}).

\begin{table}[ht]
\begin{fntable}[0.6\columnwidth]
\begin{tabular}{c|c|c|c|c}
\hline
Isotope & nat. abund. 
    & Q-value & \tzn & Experiment  \\ 
   &    (\%)    & (keV) & ($10^{20}$yrs) & \\  \hline 
 \caav &  0.187& 4272 $\pm$ 4 & 0.44$^{+0.06}_{-0.05}$ & {\small CANDLES}\\
  \gess & 7.8 & 2039.006 $\pm$ 0.050 & 15 $\pm$ 1 & GERDA, MAJORANA \\ 
  \seza & 9.2 & 2995.5 $\pm$ 1. 9 &  0.92 $\pm$ 0.07& SuperNEMO, LUCIFER\\
  $^{96}$Zr & 2.8 & 3347.7 $\pm$ 2.2 & 0.23 $\pm$ 0.02 & - \\
  $^{100}$Mo & 9.6 & 3034.40 $\pm$ 0.17 & 0.071 $\pm$ 0.004 & AMoRE \\
  $^{110}$Pd & 11.8 & 2017.85 $\pm$ 0.64 & - & -\\ 
  \cd & 7.5 & 2813.50 $\pm$ 0.13 & 0.28 $\pm$ 0.02 & COBRA, CdWO$_4$\\
  $^{124}$Sn  & 5.64 & 2287.8 $\pm$ 1.5 & - & - \\
  \tehd   & 34.5 & 2527.518 $\pm$ 0.013 & 6.8 $^{+1.2}_{-1.1}$ & CUORE \\
  \xehs & 8.9 & 2457.83 $\pm$ 0.37 & 21.1 $\pm$ 2.5 & EXO, NEXT\\ 
            & & & & KamLAND-Zen \\
  \nd & 5.6 & 3371.38 $\pm$ 0.20 & 0.082 $\pm$ 0.009 & SNO+, DCBA\\
  \hline
\end{tabular}
\end{fntable}
\caption{Table showing the eleven candidate isotopes with a Q-value larger than
2 MeV. Given are the natural abundances, Q-values as determined from precise
Penning trap measurements (those with sub keV errors) or from the Atomic Mass Evaluation 2003 \cite{aud03},
the measured averaged \zbb half lives as recommended in \protect \cite{bar10} plus the recent measurement of
\xehs  \protect \cite{ack11}. The last column shows the experiments addressing the measurement of the corresponding 
isotope. For some experiments only the ''default" isotope is mentioned as they
have the option of exploring several ones. Several additional research and development projects
are ongoing.}
\label{tab:isotopes} 
\end{table}

As mentioned, most experiments follow the approach that the source is equal to the detector, i.e. building a
detector which contains the isotope of interest. Technologies used for that are semiconductors,
cryogenic bolometers, scintillators and liquid noble gas detectors. The alternative is to use tracking
devices in form of TPCs containing thin foils of double beta emitters. Here single electron spectra
and opening angles can be measured as well.

\section{Experimental status}
Three different goals are considered for future investigation depending on the outcome 
of the individual steps. The first one is to probe the claimed
observation of \obb~ in \gess with a half-life of $2.23 \pm 0.04 \times 10^{25}$yrs \cite{kla04,kla06}.
If this is confirmed, the next generation will collect sufficient statistics for a precision
half-life measurement. It might be even possible to perform an intrinsic consistency check
by also looking at the \obb~ decay into the first excited state \cite{duerr11}. Furthermore, due to the "multi-isotope"
approach an observation in one isotope predicts a half-life for the other ones 
by taking into account the uncertainties in nuclear matrix elements, see section~(\ref{sec:nme}) . 
Observing peaks at different positions in energies within the right range of half-lives excludes potential unrecognised backgrounds.
Anyhow, the ''multi-isotope'' ansatz is needed to compensate for matrix element uncertainties.\\
In case the evidence is not confirmed, the next goal must be the region of the inverted hierarchy, i.e 
\ema below $\approx$ 50 meV. The neccessary half-life requirement to touch this region is given in \cite{dueck11}.
As the requirements on mass and background for this purpose are  already demanding, fully excluding the
IH is more than challenging.
If there is no signal found in the IH the final is the exploration of 
the normal hierarchy. However, for this ton scale experiments with extraordinary low background
have to be considered and completely new background components have to taken into account,
for example neutrino-electron scattering due to solar neutrinos \cite{bar11}. For that half-lives
well beyond 10$^{28}$ yrs have to be measured.\\

\begin{figure}[ht]
\begin{center}
\includegraphics[width=0.43\columnwidth]{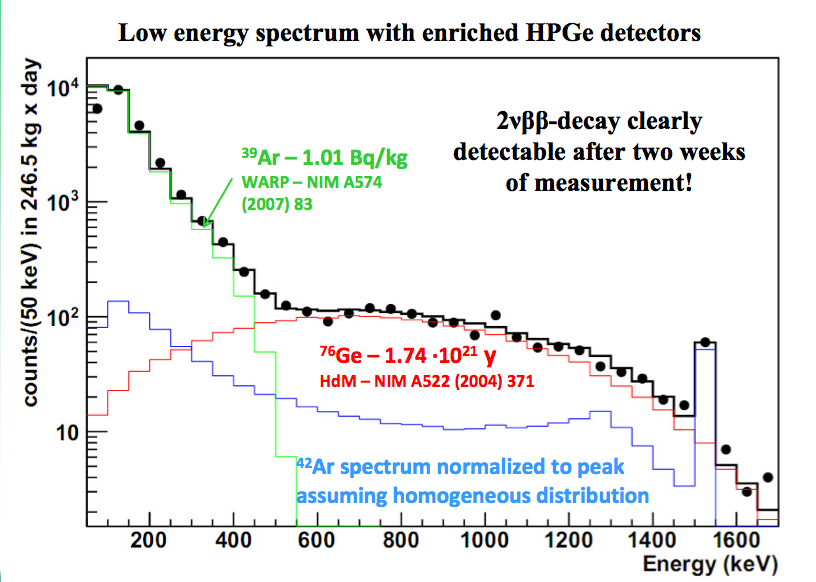}
\includegraphics[width=0.37\columnwidth]{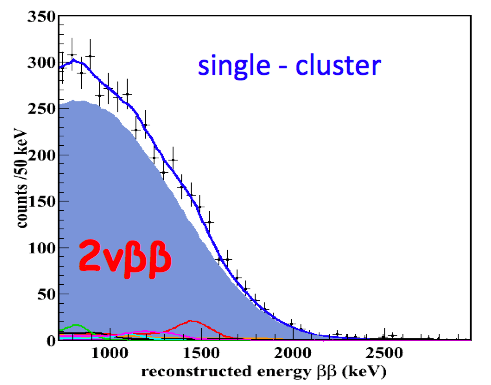}
\caption{Left: Spectrum from a commissioning run of GERDA using 3 enriched Ge-detectors. The three major visible 
components are the $^{39}$Ar - decay, a $^{42}$K contribution due to $^{42}$Ar decay and the \zbb of \gess.
Right: Initial results of EXO-200 lead to a first discovery and half-life measurement of the \zbb of \xehs being the dominant
contribution to the spectrum (from \protect \cite{gra11}, details can be found in \cite{ack11}).}
\label{fig:spectra}
\end{center}
\end{figure}

Currently the field is in the transition towards the next generation of experiments.
The year 2011 is seeing the start of four new double beta experiments, namely GERDA (using \gess), EXO and KamLAND-Zen (using \xehs) and CANDLES (using \caav). 
GERDA \cite{abt04} is a next generation experiment based on Ge-semiconductors containing \gess based in the Gran Sasso Underground
Laboratory (Italy). 
The idea is to run bare crystals within LAr which serves as shielding
and cooling. In a first phase a total of eight isotopically enriched detectors (from the former Heidelberg-Moscow and IGEX experiments) with a total mass of 17.7 kg have been deployed and official data taking started  on 1. Nov. 2011. A spectrum from a commissioning run using 3 enriched detectors is shown in Figure~(\ref{fig:spectra}). As can be
seen the \zbb contribution is already clearly visible. Currently, for a second phase another 35.5 kg of enriched Ge is in the process of conversion into BEGe detectors. These point-contact detectors allow a optimised separation of single site energy depositions (like double beta decay) and
multiple site interactions (most of the background like Compton scattering with an additional second interaction) by using pulse shape analysis
\cite{bud09}.\\
Due to the relative cheap enrichment of noble gases, two large scale experiments based on enriched \xehs have started. 
First is EXO-200, using 
about 175 kg LXe in form of a TPC located at WIPP (USA). The early data look very promising (Figure~\ref{fig:spectra}) 
and a first half-life for the \zbb of \xehs could be determined, see Table~(\ref{tab:isotopes}). 
A unique option explored for the future would be the detection
of the daughter ion which should result in major background reduction. A second Xe approach is KamLAND-Zen using the
well understood infrastructure of the KamLAND experiment. To perform double beta decay with Xe-loaded liquid scintillator
a special mini-balloon was constructed and deployed within KamLAND. In this way about 330 kg of enriched Xe could
be filled in the detector and data taking has started in September 2011.\\
Last but not least there is CANDLES using 305 kg of CaF$_2$ scintillators, focussing on \caav , the isotope with 
the highest Q-value of all double beta emitters. The experiment is installed in the Kamioka mine (Japan) and data taking has started
recently.

\section{Future experiments and ideas}
With the experiments mentioned in the previous session running, there will be more online in a few years time and additionally
still interesting research and development is ongoing, a few of them are mentioned here, see Table~(\ref{tab:isotopes}).
From the semiconductor point of view, two more projects exist, MAJORANA and COBRA.
MAJORANA is using Ge-detectors as GERDA and is considering to have up to 30 kg enriched detectors running by 2014 in the
Sanford Underground Laboratory (USA).
COBRA is exploring CdZnTe semiconductor for the search of \cd . The detectors are much smaller than the typically used Ge-diodes
thus having a much higher granularity and discrimination against multi-site events will be realised by multi-detector events.
Furthermore, for practical purposes it is convenient that these detectors will run at room temperature. A unique feature
which is explored is pixelisation which should allow massive background reduction by particle identification due to 
tracking. A prototype detector of about 0.5 kg is currently installed in the Gran Sasso Laboratory.\\
In a similar spirit as KamLAND-Zen another large scale scintillator experiment is in the building up phase.
SNO+ aims to use in a second phase of operation (a first phase should be pure scintillator mainly for the study of solar neutrinos)
 1000 t of Nd-loaded scintillator for the search of \nd decay.  
The default assumption is a 0.1\% loading of the scintillator, resulting in an amount of 43 kg of \nd , while an optimisation of 0.3\%
loading is under investigation. \nd has the second highest Q-value of all double beta isotopes (3371 keV) and was always
among the theoretically most preferred isotopes. SNO+ is supposed to start in 2013. Other groups explore various solid 
state scintillators, for example CdWO$_4$.\\
A different class of detectors not mentioned yet but with a lot of options are cryogenic bolometers. 
The most advanced approach is CUORE using 750 kg of TeO$_2$ crystals to search for the 
\tehd decay. The benefit of using Te is the high natural abundance of \tehd. This experiment
is in the building up phase at Gran Sasso Laboratory and is planned to start data taking with
the full amount in 2014. Various other bolometers are currently explored like CaMoO$_4$ (AMoRE) 
and ZnSe (LUCIFER) to search for $^{100}$Mo and \seza respectively.\\
Finally tracking devices in form of TPCs using the double beta emitter in form of thin foils are explored.
Due to the tracking it is possible to measure the individual electron energies and the opening angle
between them. This might be important if \obb is discovered as the neutrino mass mechanism
and right-handed weak current contribution differ significantly in these quantities.
As a next step in a serious of experiments SuperNEMO is planning to use at least 100 kg of \seza
in form of 20 TPC modules with 5 kg source mass per module. A first demonstrator module
is supposed to start data taking in 2014.

\section{Nuclear matrix elements - Theory and experiment}
\label{sec:nme}

\begin{figure}[ht]
\begin{center}
\includegraphics[width=0.9\columnwidth]{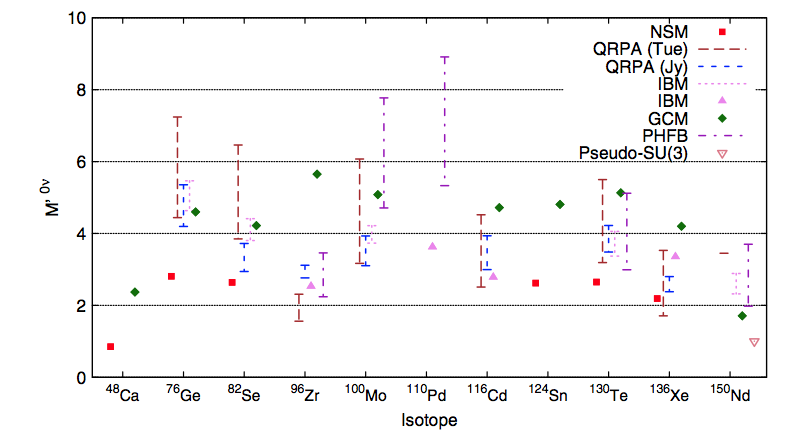}
\caption{Compilation of nuclear matrix element calculations for the eleven relevant isotopes. Existing publications
were used and normalised to a nucleon radius of $r_0 =$ 1.2 fm and $g_A$=1.25 (from \protect \cite{dueck11}).}
\label{fig:nme}
\end{center}
\end{figure}

The conversion of an observed half-life or its limit into \ema requires the knowledge of  the
nuclear transition matrix elements, see Equation~(\ref{eq:conv}). Any uncertainty in the \nme due to its
square dependence results in a significant uncertainty in the deduced neutrino mass.
This is another important argument for the multi-isotope approach. Various theoretical
methods have been applied in the past, dominantly nuclear shell model (NSM) and quasi-random
phase approximation (QRPA) calculations, which were recently joined by new methods
like the Interaction Boson Model (IBM) and energy density functional treatment (GCM). 
A compilation of the available calculations normalised to one value of r$_0$ and g$_A$ is given 
in Figure~(\ref{fig:nme}). As can be seen there is quite some spread in the obtained values and uncertainties
are not claimed for all methods.\\
 As all double beta emitters are even-even nuclei their ground
state transitions are characterised as $0^+ \ra 0^+$ transitions. Given that fact that almost all
\zbb half-lives are known the associated matrix element $ M^{2\nu}$ can be deduced. However,
these are pure Gamow-Teller transitions only mediated by the $1^+$ states of the intermediate nucleus
and restricted the less than 5 MeV and they contribute only a fraction to $ M^{0\nu}$. In the latter case
all levels up to about 100 MeV with all multipolarities can contribute. Furthermore,
effects and treatment of short range correlations become very important \cite{sim11}. 
The matrix elements itself must be calculated separately for the individual processes discussed in section (\ref{sec:intro}).
Thus, they contain information on the physics process involved as well \cite{dep07,fog09,fae11}.\\
Given the complexity of the problem and a lack of missing experimental information, a program was
started to improve the situation from the experimental point \cite{zub05}. Part of the program is
precise Q-value determinations using Penning traps (see Table~\ref{tab:isotopes}), charge exchange
reactions in form of $(^3He,t)$ and $(d,^2He)$ reactions to determine the Gamow-Teller strength $B_{GT}$
for the transitions to the intermediate $1^+$-states and nucleon transfer reactions.This lead
to new insights and refinements of the calculations.

\section{Alternative processes including positrons and electron capture}
An equivalent process to the one discussed is $\beta^+\beta^+$-decay also in combination with
electron capture (EC). There are three different variants possible depending on the Q-value:
\begin{eqnarray}
\label{eq:ecec}
(Z,A) &\ra& (Z-2,A) + 2 e^+ (+ 2 \nel) \quad \mbox{(\bpbp)}\\
e_B^- + (Z,A) &\ra& (Z-2,A) + e^+ (+ 2 \nel) \quad \mbox{(\bec)} \\
2 e_B^- + (Z,A) &\ra& (Z-2,A) (+ 2 \nel) \quad \mbox{(\ecec)}
\end{eqnarray}

\begin{figure}[ht]
\begin{center}
\includegraphics[width=0.8\columnwidth]{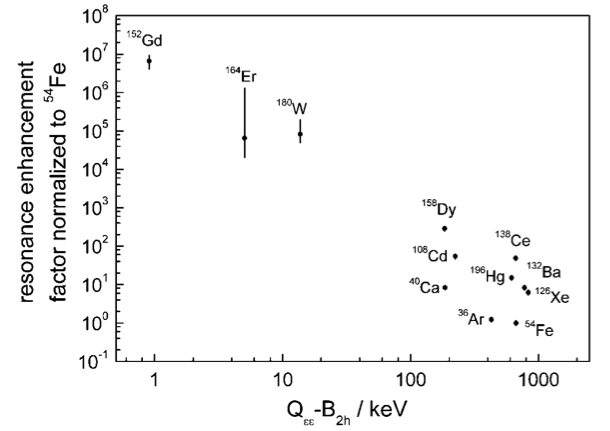}
\caption{Resonance enhancement factor for neutrinoless ECEC relative to $^{54}$Fe based
on precision mass spectrometry with Penning traps (from \protect \cite{eli11}).}
\label{fig:reso}
\end{center}
\end{figure}

$\beta^+\beta^+$ is always accompanied by EC/EC or $\beta^+$/EC-decay.
The positron production reduces the effective $Q$-value by 2$m_ec^2$ per positron. 
Therefore, the rate for $\beta^+\beta^+$ is small and energetically only
possible for six nuclides, however it would have a striking signature with four 511 keV
gamma rays. It was shown that the $\beta^+$/EC-mode has
an enhanced sensitivity to right-handed weak currents \cite{hir94} and might
be valuable to explore if \obb is discovered. The 
full Q-value is available in the  \ecec mode which is the hardest
to detect experimentally. However, it was proposed  \cite{ber83,suj04} that if an
excited state of the daughter nucleus is degenerate with the original ground
state a resonance enhancement in the decay rate could occur and the de-excitation
gammas would serve as a nice signal. Due to the sharpness of the resonance
a more detailed study of candidates had to wait for Penning traps entering the
field and exploring reasonable candidates. The most reliable one seems to 
be $^{152}$Gd (see Figure~\ref{fig:reso}) where such a scenario is realised 
\cite{eli11}. Despite this nice effect, to achieve the  same sensitivity of \ema as in
\obb seems to require a measurement an order of magnitude longer making
this method slightly less attractive.

\bibliographystyle{pramana}
\bibliography{references}

\end{document}